\documentclass[sigplan,screen]{acmart}

\AtBeginDocument{%
  \providecommand\BibTeX{{%
    \normalfont B\kern-0.5em{\scshape i\kern-0.25em b}\kern-0.8em\TeX}}}

\setcopyright{none}
\renewcommand\footnotetextcopyrightpermission[1]{}
\setcopyright{none}
\makeatletter
\renewcommand\@formatdoi[1]{\ignorespaces}
\makeatother

\begin{document}

\title{Apples and Cars: a Comparison of Security}

\author{Zhendong Ma}
\email{Zhendong.Ma@de.bosch.com}
\affiliation{%
  \institution{Bosch Engineering GmbH}
  \city{Holzkirchen}
  \country{Germany}
}

\begin{abstract}
Cybersecurity has gained importance for cars that increasingly rely on software and networks. ``Smartphone on wheels'' is often used as an analogy to highlight the need for security. As a high-value target of cyberattacks, modern smartphones implement layers of protection. Automotive embedded systems share many similarities with smartphones. We compare the security architecture of an iPhone and a car to identify gaps and discuss the potentials for the cars of the future. 
\end{abstract}

\keywords{security architecture, automotive, smartphone}

\maketitle
\pagestyle{plain}

\section{Introduction}

Modern cars are often referred to as ``smartphone on wheels'', because the embedded systems in a car are programmed and connected using similar technologies that power a smartphone in a user's pocket. Smartphones are a high-value target of cyberattacks. Consequently, layers of security measures are in place to protect the hardware, software, and data against a wide spectrum of threats. The system and threat model of a connected car have many similarities to a smartphone. Assuming that smartphones have more advanced security, questions arise as what are the commonalities and differences in securing a smartphone comparing to a connected car, and what can automotive security learn from it? In this paper, we compare the security architecture of an Apple iPhone\cite{apple21} with that of typical connected cars. The comparison shows gaps of the cars of today and potentials for the cars of tomorrow.  

\begin{figure}[h]
  \centering
  \includegraphics[width=\linewidth]{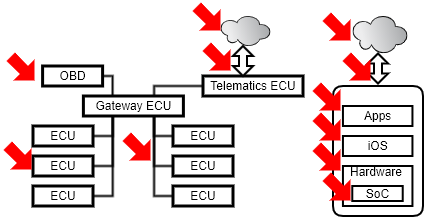}
  \caption{System and threat model of a car and an iPhone}
  \label{fig:model}
\end{figure}

Figure \ref{fig:model} illustrates the system and threat model of both systems. The red arrows indicate the attack surface. A car with a domain-based architecture uses a gateway to divide Electronic Control Units (ECU) into domains according to their functionalities. The gateway ECU routes and filters cross-domain communications. The threat model includes attacks against the cloud backend, network-based attacks via connectivity interface (i.e. the telematics ECU), local attacks via on-board OBD port and the in-vehicle communication network, and attacks on ECUs via physical access.          

An iPhone consists of hardware including a highly integrated System-on-Chip (SoC) and peripherals, the operating system iOS, and Apps from Apple and third-party. Its threat model includes attacks against the iCloud server, networked attacks with or without user interaction (the so-called ``zero-click attack''), attacks via installed Apps, attacks on iOS, and hardware attacks via physical access down to the silicon.

\section{Comparison of security}

\textbf{Hardware security}
iPhone's threat model assumes that an attacker has physical access to the internals of the SoC to intercept, extract and manipulate the data on the bus and in the random-access memory (RAM) and non-volatile memory (NVM). Inside the Apple SoC there is a coprocessor named \emph{Secure Enclave} with its own microkernel operating system for cryptographic operation. It isolates cryptographic secrets and operation from the application processor and the main memory and protects the cryptographic module from side-channel attacks. The Secure Enclave is equipped with tamper detection and power and clock monitoring circuit to protect itself from tampering and dynamic power analysis. Both the Secure Enclave and the application processor have its hardware-based inline AES engine, which encrypts data written to memory. In addition to encrypted memory, Secure Enclave also implements measures for memory authentication and replay protection to ensure memory security within the SoC. The build-in public key accelerator supports RSA and ECC encryption and signing. An immutable Boot ROM provides a hardware root-of-trust for system boot.            
 
The ECUs inside the cars typically use microcontrollers ($\mu$C) for time-critical functions such as steering or engine control, and microprocessors ($\mu$P) for computation intensive functions such as autonomous driving (AD) or in-vehicle infotainment. The state-of-the-art $\mu$Cs are equipped with Hardware Security Modules (HSM) for secure key storage and cryptographic operation. A HSM has its own processor and RAM for a physically isolated execution environment. It communicates with the main CPU via the system bus. It has hardware accelerator for AES. Some models include hardware accelerator for asymmetric algorithm such as RSA and ECC. Hardware security of $\mu$P is typically achieved by trusted execution environment (TEE) technology, which is a virtual execution environment created on the main processor and supported by hardware-based access control.      

In comparison, Secure Enclave is embedded in the SoC package. HSM is located outside the main processors of the $\mu$C. TEE is a software solution supported by hardware. The more a secure execution environment can be isolated, the higher the security level it will achieve.       

\textbf{System and software security}
iPhone's system security ensures that the device runs only authentic software from Apple and the system maintains its integrity during runtime. Secure boot ensures that the software in each stage verifies the integrity of the next stage before handing over control. 
The iBoot bootloader is a security-enhanced build that hardens the software against typical C vulnerabilities such as buffer-overflow, heap exploitation, and type confusion. iOS uses several integrity protection mechanisms. \emph{Kernel integrity protection} creates a protected physical memory region for the kernel and its extensions. The region is locked after boot. \emph{Fast permission restrictions} use a CPU register to restrict memory read/write/execute permissions on a per-thread basis. During boot, iBoot loads coprocessor's firmware into protected memory region and the system prevents executable and writeable mapping on that region. \emph{Pointer authentication codes} protect program control flows by preventing modification of function pointers and return addresses. All executable code are signed and verified before execution in the memory. \emph{Page protection layer} ensures that only it can alter memory pages containing user code such that an attacker cannot tamper the code after signature verification. Third-party apps are sandboxed to restrict their access to other files and system resources. Further apps memory protections include Address Space Layout Randomization (ASLR) and Execute Never (XN) which prevent code execution from certain memory pages. Apple over-the-air (OTA) update process includes additional parameters to personalize the update package for each device and to prevent replay and downgrade attack. To request an update, an iPhone sends a nonce and its unique device ID to the server. The server signs the update including the installation bundle, the nonce, and the device ID.   

In automotive domain, secure boot gradually becomes a standard security feature to ensure software authenticity and integrity inside ECUs. Although some $\mu$Cs include memory protection unit that restricts read/write access to certain memory ranges based on the initiator's ID, it is rather a safety feature. \emph{Runtime Manipulation Detection} takes a step further towards the protection of the integrity of runtime memory. However, currently measures mainly rely on reducing vulnerabilities in ECU software by emphasizing secure coding practice (e.g. MISRA C/C++ coding standard) and extensive testing (e.g. fuzzing). Many ECUs can be flashed with signed software. However, legacy ECUs or those with limited capacities often rely on a tester or an OTA master for signatures verification instead of doing it inside the target ECU, which increases the risk of exploitation of race conditions.                 

\textbf{Network security}
iPhones and cars use very similar network security technologies for secure communication. TLS, VPN, or the latest Wi-Fi or Bluetooth security standards are commonly implemented in both worlds. Apple randomizes the addresses and identifiers in wireless communication to prevent device tracking. It uses application firewall to filter communication on a per-application basis. Network security measures in cars are often allocated at entry point. For in-vehicle network, commonly a part of the Ethernet is secured by TLS and a handful of bus signals are protected by Secure Onboard Communication (SecOC) for integrity and freshness. A gateway ECU implements firewall and packet/message filtering on cross-domain communications.           

\section{Conclusion and discussion}

The comparison shows that the ``Apple'' is several years ahead of the ``cars''. Despite being in different product categories, the security model of an iPhone from today should be the goal of the cars of tomorrow. The security of iPhone characterizes a zero trust model from the silicon level up, while cars still follow a network-based perimeter model. As the car industry evolves towards high performance vehicle computers and zone architecture, the security model needs to be adapted to couple with increased system integration and function mix. iPhones demonstrate that it is technologically feasible to reach a high level of security without sacrificing performance and usability. Despite the effort, iPhone security measures are still bypassed or defeated on a rare but constant basis. Some of the attack methods will be applicable to automotive embedded systems. Recognizing the evolving threat model will give the automotive industry incentive for raising the bar on security for the cars of tomorrow.              

\bibliographystyle{ACM-Reference-Format}

\end{document}